\documentclass[conference]{IEEEtran}
\IEEEoverridecommandlockouts
\usepackage[english]{babel}
\usepackage[utf8x]{inputenc}
\usepackage{cite}
\usepackage{amsmath,amssymb,amsfonts}
\usepackage{amsthm}
\usepackage{algorithmic}
\usepackage{algorithm}
\usepackage{graphicx}
\usepackage{textcomp}
\usepackage{xcolor}
\usepackage{stfloats}

\def\BibTeX{{\rm B\kern-.05em{\sc i\kern-.025em b}\kern-.08em
		T\kern-.1667em\lower.7ex\hbox{E}\kern-.125emX}}

\begin{document}
	
\title{
	Caching as an Image Characterization Problem using Deep Convolutional Neural Networks
\thanks{This is an accepted version of ICC2020 \textcopyright2020 IEEE. Personal use of this material is permitted. Permission from IEEE must be obtained for all other uses, in any current or future media, including reprinting/republishing this material for advertising or promotional purposes, creating new collective works, for resale or redistribution to servers or lists, or reuse of any copyrighted component of this work in other works.}
}

\author{\IEEEauthorblockN{Yantong Wang}
	\IEEEauthorblockA{Center for Telecommunications Research\\
		Department of Engineering, King's College London\\
		London WC2R 2LS, U.K. \\
		yantong.wang@kcl.ac.uk}
	\and
	\IEEEauthorblockN{Vasilis Friderikos}
	\IEEEauthorblockA{Center for Telecommunications Research\\
		Department of Engineering, King's College London\\
		London WC2R 2LS, U.K. \\
		vasilis.friderikos@kcl.ac.uk}
}

\maketitle

\begin{abstract}
	Caching of popular content closer to the mobile
user can significantly increase overall user experience as well as
network efficiency by decongesting backbone network segments
in the case of congestion episodes. In order to find the optimal
caching locations, many conventional approaches rely on solving	a complex optimization problem that suffers from the curse of
dimensionality, which may fail to support online decision making.
In this paper we propose a framework to amalgamate model
based optimization with data driven techniques by transforming
an optimization problem to a grayscale image and train a convolutional neural network (CNN) to predict optimal caching location policies. The rationale for the proposed modelling comes
from CNN’s superiority to capture features in grayscale images
reaching human level performance in image recognition problems. The CNN is trained with optimal solutions and numerical
investigations reveal that the performance can increase by more
than 400\% compared to powerful randomized greedy algorithms. To this end, the proposed technique seems as a promising way
forward to the holy grail aspect in resource orchestration which
is providing high quality decision making in real time.
\end{abstract}


\section{Introduction}
\label{sec:intro}

The proliferation of popular content on the Internet is
creating a massive amount of aggregate data that need to be
transported across, mostly, congested links. Bringing popular
content closer to the end users via caching, in order to ease the heavy burden of transporting such data through the backhaul and fronthaul of mobile network and at the same time increase
user experience, has received significant research attention
over the last few years \cite{SurveyCaching}. 
Compared with other caching policies, proactive caching makes a trade-off between storage and transmission. 
Generally, there are two inter-linked subproblems in proactive caching that need to be tackled; firstly,
the issue of where to cache and secondly what to cache in each
selected location. In this paper, we focus mainly on the first problem which relate on choosing the edge clouds for hosting
popular content. 
For the purpose of deriving optimal decision making for content placement a popular approach is based on
using mixed integer linear programming (MILP) formulations. Typically such formulations consider both content hosting cost at a specific edge cloud and transmission cost which depends on the topological proximity between the requested content
and the end user, with constraint of storage space, limitation of bandwidth and the requirements for Quality of Services(QoS) \cite{wang2019proactive}. However, these type of problems belong to the family of $\mathcal{NP}$-hard\cite{zappone2019wireless} optimization problems. Therefore, due to the
curse of dimensionality such model-based methods cannot be deemed as suitable to support real-time decision making in 5G
and beyond mobile communication networks.


Recently, deep learning (DL) technology, a member of the
broader family of machine learning, attracted significant attention from both academia and industry. DL is based on artificial
neural networks but has multi hidden layers in between the input and output layer and the applications are not only limited
to speech or image recognition, but also expended to a diverse
set of domains such as self-driving cars, medical diagnosis and
playing games such as Go\cite{sze2017efficient}.  DL is also gaining significant
attention in network research and already DL techniques have been used for resource management, routing and other network related operations \cite{sun2019application}. Undeniably, these AI-based technologies, like DL, will play an important role in future network design and operation. The reason comes from two main aspects: i) the explosive growth of traffic data in future network provides large datasets for DL training\cite{cheng2017mobile}; ii) the improvement of hardware allows for practical implementation AI-based approaches, for example, the graphics processing units (GPUs) execute DL at orders of magnitude faster than traditional Central Processing Units (CPUs). Compared to some other deep learning architectures, Convolutional Neural Network (CNN) has the feature of weight-sharing, which means the same set of weights are used in the processing. Additionally, recent results show that a CNN can exceed
human-level accuracy in image recognition \cite{sze2017efficient}. 

\begin{figure}[htb]
	\centering
	\includegraphics[width=0.48\textwidth]{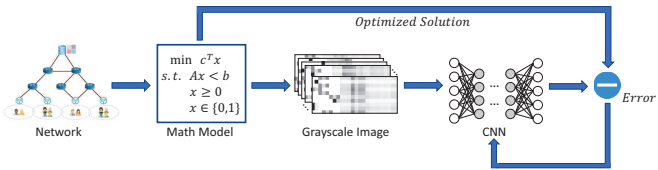}
	\caption{Training Process}
	\label{fig:training}
\end{figure}

\begin{figure}[htb]
	\centering
	\includegraphics[width=0.48\textwidth]{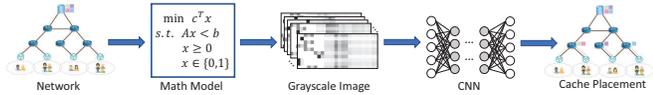}
	\caption{Testing Process}
	\label{fig:testing}
\end{figure}

In this paper, we transform the original optimization problem into a grayscale image in order to train a deep CNN as shown in Figure \ref{fig:training}. Figure \ref{fig:testing} presents the process of performance testing. The details of processing steps will be discussed in the following sections. We note that in this work we aim to explore spatial information of the requests for popular content and based on optimal allocation learn caching locations from derived grayscale images. For real time
decision making some form of greedy heuristics need to be
used. The main advantage of the proposed approach is that
it can provide real time decision making and the quality of
those decisions can be more than 400\% better compared to a randomized greedy heuristic. To this end, the contributions can be summarized as follows:
\begin{itemize}
\item We explicitly factor the optimization problem into the grayscale image therefore we can benefit from CNN's superiority to capture spatial features as being done for image classification/recognition.
\item In the proposed approach time consuming optimal policies can be computed offline together with the training of
the deep CNN which then can provide real time decision
making. 
\end{itemize}

\section{Related Work}
\label{sec:related}
The existing literature has studied a plethora of approaches related to the popular content placement problem. Conventionally, caching assignment problem is modelled as an optimization problem which is then solved by convex optimization \cite{yang2018cache}, integer linear programming \cite{zheng2016optimal}, \cite{wang2019proactive} or game theoretic methodologies \cite{fang2015energy}. Recently, deep neural network (DNN) has been used as an important tool for solving caching problems \cite{chen2019artificial}. There are already a number of research works on combining machine learning with 'what to cache'. The work in \cite{tanzil2017adaptive} use an extreme-learning machine neural network to estimate the popularity of content. In \cite{wei2018joint}, DNN is employed for cache replacement and content delivery. The authors in \cite{chen2017caching} predict the user mobility and content request distribution via echo-state networks. In \cite{lei2019deep}, a deep learning network is proposed to predict the data package popularity based on the user request packages. However, there are very limited work solving 'where to cache' problem via deep learning. A DNN is trained to solve a linear programming problem in \cite{lei2017deep}. In \cite{lee2018deep}, the authors apply DNN to solve linear sum assignment problem. Generally, the cache assignment problem is modelled as an integer linear programming model, which is $\mathcal{NP}$-hard while linear programming is in $\mathcal{P}$, and the constrains in caching placement problem are more complicated than linear sum assignment problem. In this paper, we transform the original MILP model into a grayscale image as the input, then use a CNN which is trained with optimal solutions to decide the cache assignment.

\section{System Model}
\label{sec:model}

The network is modelled as an undirected graph $\mathcal{G}=\{\mathcal{V},\mathcal{L}\}$, where $\mathcal{V}$ represents the set of vertices and $\mathcal{L}$ denotes the set of links. We define a set $\mathcal{E} \subseteq \mathcal{V}$ that consist of the edge clouds (ECs), i.e., content routers which can be selected as information hoster\footnote{The term \textit{edge cloud} and \textit{content router} are used interchangeably in the rest of the paper}. By $\mathcal{A} \subseteq \mathcal{V}$, we define the set of access routers (ARs) that mobile users might move due to their mobility; this information is assumed to be accurately known using historical data that an operator can explore. We note that in the general case it is possible that the following holds $\mathcal{E} \cap \mathcal{A} \neq \emptyset$, which means some access routers have the ability to host the content.

For network modelling reasons and without loss of generality, we assume each mobile user is associated with a single request flow and $k\in\mathcal{K}$ is defined as the set of flows traverse the network. Each flow $k$ has the following three associated properties: $s_k$, which is the size of cache items for flow $k$; $b_k$, the required bandwidth of flow $k$ and $p_{ka}$ which encapsulate the probability for flow $k$ to move to AR $a$, where $a\in \mathcal{A}$. Similarly, for each potential EC $e \in\mathcal{E}$: with $w_e$ we express the remaining cache space in EC $e$ and for each link $l\in\mathcal{L}$ we have the available capacity $c_l$. 




Furthermore, we use $N_{ae}$ to represent the number of hops from AR $a$ to EC $e$ which follows the shortest path; with $N^T$ we express the number of hops from AR to data center; $\alpha$ is the weight measuring cache cost and $\beta$ is the weight of transmission cost. The key notations we used in this paper are summarized in Table \ref{tab:Notations}.

\begin{table}[htb] 
	\centering 
	\caption{Summary of Main Notations}
	\begin{tabular}{c|l} 
		\hline
		\hline
		$\mathcal{K}$ & the set of flows/requests \\
		$\mathcal{L}$ & the set of links \\
		$\mathcal{A}$ & the set of ARs \\
		$\mathcal{E}$ & the set of ECs \\
		\hline
		$\alpha$ & the weight of caching cost\\
		$\beta$ & the weight of transmission cost\\
		$N_{ae}$ & number of hops from AR to EC, depends on\\ & \textbf{network topology}\\
		$B_{lae}$ & relationship between link and paths, depends on \\ & \textbf{network topology}\\
		$N^T$ & number of hops from AR to data center, depends on \\ & \textbf{network topology} \\
		\hline
		$p_{ka}$ & user moving probability, depends on \textbf{user behaviour} \\
		$s_k$ & interested content storage space requirement, depends on \\ & \textbf{user behaviour} \\
		$b_k$ & required bandwidth, depends on the \textbf{user behaviour} \\
		$w_e$ & available space, depends on \textbf{network resource} \\
		$c_l$ & available link capacity, depends on \textbf{network resource} \\
		\hline
		$|\mathcal{X}|$ & the cardinality of set $\mathcal{X}$ \\
		$P'$ & the transpose of matrix $P$\\
		\hline
		$x_{ke}$ & binary decision variable, which indicates the assignments of \\ & flow and edge cloud \\
		$y_{kl}$ & binary decision variable, which expresses the usage of flow \\ & and link \\
		$z_{kae}$ & binary decision variable, which represents the path of flow \\
		\hline
	\end{tabular}
	\label{tab:Notations}
\end{table}

Based on the aforementioned network setting detailed in the previous section and in order to provide a mathematical programming framework we define the following binary decision variables,

\hangafter 1
\hangindent 1em
\noindent
\\$  x_{ke}=
\begin{cases}
1,  &\text{if content for flow $k$ is cached at EC $e$}  \\
0,  &\text{otherwise}
\end{cases}$
\\
\\$ y_{kl}=
\begin{cases}
1, &\text{if flow $k$ passes link $l$} \\
0, &\text{otherwise}
\end{cases}$
\\
\\$ z_{kae}=
\begin{cases}
1, &\text{if flow $k$ connect with $a$ and retrieve the}\\ &\text{cached content from EC $e$}\\
0, &\text{otherwise}
\end{cases}$
\\

We define the total cost ($TC$) as 
\begin{equation}
TC=\alpha\cdot C^C+\beta\cdot C^T
\end{equation}

where $\alpha$ and $\beta$ are impact factors indicating the weight of two prices. Specifically, $C^C$ is the cost of caching content, which can be written as follows \cite{vasilakos2012proactive}:
\begin{equation}
C^C=\sum_{k\in\mathcal{K}}\sum_{e\in\mathcal{E}}\frac{x_{ke}}{1-U_e}
\end{equation}
and $U_e$ is the space utilization level at EC $e$, whose value depends on different cache assignment policy:
\begin{equation}
U_e=\frac{\sum_{k\in\mathcal{K}}s_k\!\cdot\! x_{ke}}{w_e}=\sum_{k\in\mathcal{K}}q_{ke}\!\cdot\!x_{ke},\forall e\in\mathcal{E}
\end{equation}
Here we introduce $q_{ke}=s_k/w_e, \forall k\!\in\!\mathcal{K}, e\!\in\!\mathcal{E}$, which indicates the effect of caching selection on space utilization level. Then $C^C$ becomes
\begin{equation}
\label{fml:cc}
C^C=\sum_{k\in\mathcal{K}}\sum_{e\in\mathcal{E}}\frac{x_{ke}}{1-\sum_{k\in\mathcal{K}}q_{ke}x_{ke}}
\end{equation}
It is worth noting that the denominator in formula \eqref{fml:cc} contains the decision variable $x_{ke}$, which is non-linear. By using the trick introduced in \cite{wang2019proactive}, we can linearize it. An auxiliary decision variable is defined below:
\begin{equation}
t_e=\frac{1}{1-\sum_{k\in\mathcal{K}}q_{ke}\!\cdot\!x_{ke}},\forall e\in\mathcal{E}
\end{equation}
This definition is equivalent to the following constraints:
\begin{subequations}
\begin{align}
\label{fml:t1}
&t_e-\sum_{k\in\mathcal{K}}q_{ke}\!\cdot\!x_{ke}t_e=1,\forall e\in\mathcal{E} \\
\label{fml:t2}
&t_e>0, \forall e\in\mathcal{E}
\end{align}	
\end{subequations}
There is a product of two decision variables in equation \eqref{fml:t1}, so we rewrite the formula in terms of $\chi_{ke}$:
\begin{equation}
\chi_{ke}=t_e\!\cdot\!x_{ke}=
\begin{cases}
t_e,&\text{if $x_{ke}$ is 1}\\
0,&\text{otherwise}
\end{cases}
\end{equation}
Then $C^C$ becomes
\begin{equation}
C^C=\sum_{k\in\mathcal{K}}\sum_{e\in\mathcal{E}}\chi_{ke}
\end{equation}

$C^T$ expresses the cost for transmission, which is consisted by caching-hit and miss branch:
\begin{equation}
\begin{aligned}
C^T=C^H+C^M=\sum_{k\in\mathcal{K}}\sum_{a\in\mathcal{A}}\sum_{e\in\mathcal{E}} p_{ka}N_{ae}z_{kae}\\+\sum_{k\in\mathcal{K}}(1\!-\!\sum_{a\in\mathcal{A}}\sum_{e\in\mathcal{E}}p_{ka}z_{kae})N^T
\end{aligned}
\end{equation}
Where $C^H$ represents the cache hitting cost when the mobile user (i.e. flow $k$) connects to the AR $a$ and retrieve to EC $e$. $N_{ae}$ is the number of hops on the shortest path between connected AR $a$ and cache hosting EC $e$. Notably $N_{ae}=0$ iff $a=e$. $C^M$ is the cost for flow $k$ missing cache, which means we cannot retrieve the request from flow $k$ to a caching hoster $e$. Then the request will be redirected to the data center. Here we use $N^T$ to represent the number of hops from AR $a$ to central data server. In the current network, the request traverses an average of $10$ to $15$ hops between source and destination \cite{van2014performance}. So the value of $N^T$ is between $10$ and $15$. 

The objective of this paper is to determine the optimal caching strategy that minimizes total cost which can be formulated as:
\begin{subequations} \label{LP:main}
	\begin{align}
	\label{obj}
	&\mathop{\min}_{\substack{x_{ke},y_{kl},z_{kae}}}\; TC \\
	\textrm{s.t.}\quad
	\label{LP:con1}
	& \sum_{e\in\mathcal{E}} x_{ke}\!\leq\!1, \forall k\!\in\!\mathcal{K} \\
	\label{LP:con2}
	& \sum_{k\in\mathcal{K}} s_k\!\cdot\!x_{ke}\!\leq\!w_e, \forall e\!\in\!\mathcal{E}\\
	\label{LP:con3}
	& \sum_{e\in\mathcal{E}} z_{kae}\!\leq\!1, \forall k\!\in\!\mathcal{K},a\!\in\!\mathcal{A} \\
	\label{LP:con4}
	& z_{kae}\!\leq\!x_{ke}, \forall k\!\in\!\mathcal{K},a\!\in\!\mathcal{A},e\!\in\!\mathcal{E}\\
	\label{LP:con5}
	& \sum_{k\in\mathcal{K}} b_k\!\cdot\!y_{kl}\!\leq\!c_l,\forall l\in\mathcal{L} \\
	\label{LP:con6}
	& y_{kl}\!\leq\!\sum_{a\in\mathcal{A}}\sum_{e\in\mathcal{E}} B_{lae}\!\cdot\!z_{kae}, \forall k\!\in\!\mathcal{K},l\!\in\!\mathcal{L} \\
	\label{LP:con7}
	& M\!\cdot\!y_{kl}\!\geq\!\sum_{a\in\mathcal{A}}\sum_{e\in\mathcal{E}} B_{lae}\!\cdot\!z_{kae}, \forall k\!\in\!\mathcal{K},l\!\in\!\mathcal{L}\\
	\label{LP:con8}
	& t_e\!-\!\sum_{k\in\mathcal{K}}q_{ke}\!\chi_{ke}\!=\!1,\forall e\!\in\!\mathcal{E}\\
	\label{LP:con9}
	& \chi_{ke}\!\leq\!t_e,\forall k\!\in\!\mathcal{K},e\!\in\!\mathcal{E}\\
	\label{LP:con10}
	& \chi_{ke}\!\leq\!M\!\cdot\!x_{ke},\forall k\!\in\!\mathcal{K},e\!\in\!\mathcal{E}\\
	\label{LP:con11}
	& \chi_{ke}\!\geq\!M\!\cdot\!(x_{ke}-1)\!+\!t_e,\forall k\!\in\!\mathcal{K},e\!\in\!\mathcal{E}\\
	& x_{ke},y_{kl},z_{kae}\!\in\!\{0,1\},\forall k\!\in\!\mathcal{K},l\!\in\!\mathcal{L},a\!\in\!\mathcal{A},e\!\in\!\mathcal{E}\\
	& t_e>0,\forall e\in\mathcal{E}
	\end{align}
\end{subequations}
where $M$ is a sufficiently large number. Constraint \eqref{LP:con1} limits the number of caching ECs for each flow. \eqref{LP:con2} shows the cache capacity for individual EC, where \eqref{LP:con3}, \eqref{LP:con4} enforce the redirected path is unique and the redirected EC should host related contents. \eqref{LP:con5} imposes the bandwidth for each link. Moreover, the next two constraints \eqref{LP:con6} and \eqref{LP:con7} enforce the link which flow passed should belong to a retrieved path, and vice versa, where $B_{lae}$ shows the relationship between link and path, which could be generated from network topology by defining $$B_{lae}=
\begin{cases}
1, &\text{if link $l$ in shortest path between $a$ and $e$} \\
0, &\text{otherwise}
\end{cases}$$
The constraint \eqref{LP:con8} as well as \eqref{LP:con9}$\sim$\eqref{LP:con11} represent definition of auxiliary variable $t_e$ and $\chi_{ke}$ respectively.

\section{Deep Learning for Caching}
\label{sec:DNN}
\subsection{Mathematical Model to Grayscale Image (M2GI)}
In this part, we are going to transform the mathematical model into a grayscale image as the input of CNN. In other words, we want to construct a parameter matrix which contains maximum information of optimization model \eqref{LP:main}. One idea is putting all the parameters as listed in Table \ref{tab:Notations} together as the matrix. However, some of them are constant, such as $N_{ae}$ indicating the number of hops which relies on the network topology, while the topology is fixed in most cases once the network is constructed. Therefore these kinds of parameters have very limited contribution to distinguish different assignments since they are unchanging as the redundant information in the input dataset. By analysing the optimal solutions, we observe that there is a pattern among user behaviour, network resource and the allocation of content caching. For example, those edge clouds, which have more available hosting space, enough link bandwidth through the potential path and closer to the crowded access routers, are more likely to be selected as information hoster. So we extract those parameters depending on network resource or user behaviour in Table \ref{tab:Notations}, which are moving probability $p_{ka}$, required storage space $s_k$, required link bandwidth $b_k$, available space in edge cloud $w_e$ and available capacity in link $c_l$.

\begin{figure}[htb]
	\centering
	\includegraphics[width=0.48\textwidth]{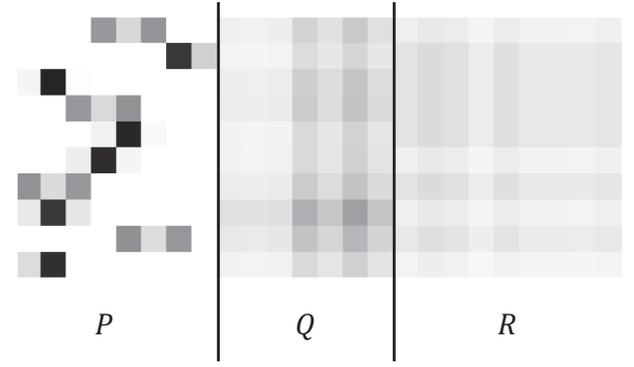}
	\caption{Feature Encoding to Grayscale Image}
	\label{fig:encoding}
\end{figure}

In Section \ref{sec:model}, $q_{ke}$ is introduced representing the effect of caching selection on storage space utilization level, which is the combination of $s_k$ and $w_e$. Similarity, we can define $r_{kl}=b_k/c_l, \forall k\!\in\!\mathcal{K},l\!\in\!\mathcal{L}$ as the impact on link bandwidth utilization level. Thus, we use the following matrix consisting of $p_{ka}$, $q_{ke}$ and $r_{kl}$ as the feature to be learnt and predicted by CNN:
$$
\setlength{\arraycolsep}{1pt}
\left[
\begin{array}{ccc|ccc|ccc}
p_{11}&\cdots&p_{1|\mathcal{A}|}&q_{11}&\cdots&q_{1|\mathcal{E}|}&r_{11}&\cdots&r_{1|\mathcal{L}|}\\
p_{21}&\cdots&p_{2|\mathcal{A}|}&q_{21}&\cdots&q_{2|\mathcal{E}|}&r_{21}&\cdots&r_{2|\mathcal{L}|}\\
\vdots&\ddots&\vdots&\vdots&\ddots&\vdots&\vdots&\ddots&\vdots\\
p_{|\mathcal{K}|1}&\cdots&p_{|\mathcal{K}||\mathcal{A}|}&q_{|\mathcal{K}|1}&\cdots&q_{|\mathcal{K}||\mathcal{E}|}&r_{|\mathcal{K}|1}&\cdots&r_{|\mathcal{K}||\mathcal{L}|}\\
\end{array}
\right]
$$
Then we can build the grayscale image based on the above matrix after normalization. Figure \ref{fig:encoding} shows a demonstration where $|\mathcal{K}|=10$, $|\mathcal{A}|=8$, $|\mathcal{E}|=7$ and $|\mathcal{L}|=9$. The darker the pixel, the larger element value in matrix.

Using the grayscale image as input and decision variable $x_{ke}$ as output, we can train the neural network. The reason why we disregard decision variable $y_{kl}$ and $z_{kae}$ is that they can be represented by $x_{ke}$. For example, from constraint \eqref{LP:con4} we get $x_{ke}$ is the upper-bound of $z_{kae}$. In most cases, the penalty of cache missing is larger than the transmission cost. So we only need to take some potential ARs (i.e. $p_{ka}$ exceeds a threshold which can be defined as the ratio between cost for caching and gain with caching \cite{vasilakos2012proactive}) into consideration. Similarly, constraints \eqref{LP:con6} and \eqref{LP:con7} limit the upper- and lower-bound of $y_{kl}$. Once $x_{ke}$ is determined, the other two are solved. 

\subsection{CNN Architecture}
\begin{figure}[htb]
	\centering
	\includegraphics[width=0.48\textwidth]{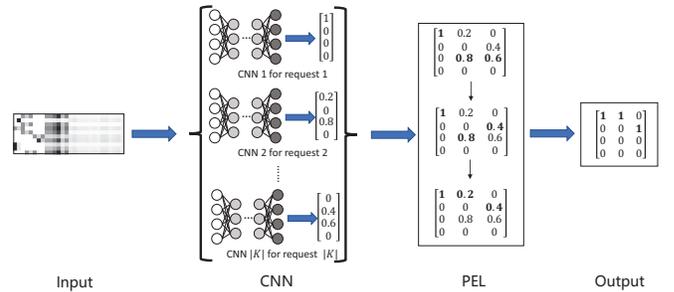}
	\caption{Proposed Algorithm Architecture}
	\label{fig:structure}
\end{figure}

As above-mentioned, the decision variable $x_{ke}$ is used as the output. Because $x_{ke}$ is a binary variable and edge clouds can be viewed as labels, we deem it a multi-label classification problem. By applying the first order strategy \cite{zhang2013review}, the original problem can be decomposed into a number of independent classification sub-problems. Each sub-problem corresponds to a related CNN. Nevertheless, the results might be sub-optimal due to the unawareness of flow correlations. Hence in \cite{lee2018deep}, a greedy collision avoidance rule is introduced after the output of DNN. In our work, we add a  performance enhancement layer (PEL) behind the output layer to improve the system performance. The architectures of our proposed learning algorithm is illustrated in Figure \ref{fig:structure}. Since each CNN is independent, training them in parallel benefits the computation complexity.

\begin{figure}[htb]
	\centering
	\includegraphics[width=0.48\textwidth]{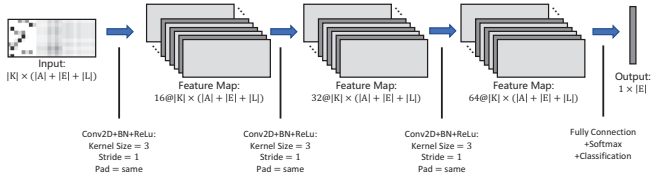}
	\caption{Designed CNN Architecture}
	\label{fig:cnn}
\end{figure}

Figure \ref{fig:cnn} shows the structure of CNN responding to user request which is constructed by following layers:
\begin{itemize}
	\item [1)] 
	Input Layer: this layer specifies the image size and applies data normalization. The image size is represented by a vector which correspond to the height, width and number of channels respectively. In this case, our input is a grayscale image, so the channel size is 1. The height is $|\mathcal{K}|$ and width is $|\mathcal{A}|+|\mathcal{E}|+|\mathcal{L}|$.
	\item [2)]
	Convolutional Layer: this layer applies sliding convolutional filters to the input. In this paper, we use three identical convolutional layers with 16, 32 and 64 filters respectively of size $3\times3$. The step size for moving the input vertically and horizontally, i.e. stride, is 1. By setting pad same, we keep the output has the same size as the input.
	\item [3)]
	Batch Normalization Layer: this layer normalize the input to speed up network training and reduce the sensitivity to network initialization.
	\item [4)]
	ReLu Layer: in this layer we use the most common activation function, rectified linear unit (ReLu), as the threshold to each element of the input. This operation is set any value less than 0 to 0 and keep the positive value. 
	\item [5)]
	Fully Connected Layer: as the name indicates, the neurons in this layer are connected to all the other neurons in the preceding layer. In our case, this layer combines the features of grayscale image to select the edge cloud for caching.
	\item [6)]
	Softmax Layer: this layer normalizes the output of Fully Connected Layer. The output of Softmax Layer consists of positive numbers whose summary is 1, which can be used as classification probabilities.
	\item [7)]
	Classification Layer: this layer uses the output of Softmax Layer for each grayscale image to assign it to one of the potential edge clouds then compute the loss.
\end{itemize}

\subsection{Performance Enhancement Layer (PEL)}
As the aforementioned, combining the output of each CNN directly may cause collision since it does not consider the nternal effect of decision variables. In this subchapter, a PEL is proposed in order to avoid the conflict. We notice that in Figure \ref{fig:structure} the result in each CNN is a $1\times|\mathcal{K}|$ one-hot vector (i.e. only a single 1 and all others 0), which depends on the output of softmax layer. The softmax layer's output is classification probability, which can be viewed as the confidence of the prediction. Considering CNN has advanced learning skills, it is a small-probability event that the optimal solution locates outside the selected assignment. Additionally, a piecewise function is added in total cost $TC$ as such penalty cost to measure the impact of invalid constraints \eqref{LP:con2} and \eqref{LP:con5}:
\begin{equation}
\label{fml:new_cost}
\begin{aligned}
TC^N=&TC\!+\!\gamma\!\cdot\!\max\bigg(0,\big(\sum_{e\in\mathcal{E}}\sum_{k\in\mathcal{K}}q_{ke}\!\cdot\!x_{ke}\!-\!1\big)\\
&\!+\!\big(\sum_{l\in\mathcal{L}}\sum_{k\in\mathcal{K}}r_{kl}\!\cdot\!y_{kl}\!-\!1\big)\bigg)
\end{aligned}
\end{equation}
where $\gamma$ is the penalty factor for invalided constraints. In PEL, we do exploration based on the predicted probability. The assignment with highest value in each row can be obtained as the initial solution. Once a new assignment with less cost $TC^N$ is found, the solution will be updated accordingly. The basic idea of PEL is viewing the initial combination of each CNN as the starting point, then searching alongside the predicted probability.  To reduce the number of attempted assignments, a threshold is set for the predicted probability, which means we only enumerate the element over that threshold. More details are illustrated in Algorithm \ref{alg:PEL} below.

\begin{algorithm}
	\caption{Performance Enhancement Layer}
	\label{alg:PEL}
	\begin{algorithmic}[1]
		\REQUIRE ~~ \\
		Grayscale image $[P,Q,R]$;\\
		Predicted conditional probability $O$; \\
		Threshold of probability $\delta$;\\
		Cost weight $\alpha$, $\beta$ and penalty factor $\gamma$; \\
		Number of hops for cache hitting $N_{ae}$ and missing $N^T$; \\
		Relationship matrix between link and path $B_{lae}$
		\ENSURE ~~ \\
		Flow assignment $X=(x_{ke})$
		\STATE Construct assignment matrix $X\to\max(0,O-\delta)$ 
		\STATE Add the maximum value of each column in $X$ to queue $\Omega$, while the rest non-zero element to queue $\Psi$
		\STATE Extract elements in $\Omega$ as initial binary assignment, determine $z_{kae}$ and $y_{kl}$ and calculate 
		cost $TC_I^N$ via \eqref{fml:new_cost}
		\FOR{Each element in $\Psi$}
		\STATE Renew assignment by using the largest element in $\Psi$ to replace the according one in $\Omega$ and update $z_{kae}$ and $y_{kl}$
		\STATE Recalculate total cost $TC_U^N$ accordingly using \eqref{fml:new_cost}
		\IF{$TC_U^N<TC_I^N$}
		\STATE Enqueue the selected element of $\Psi$ in $\Omega$, and dequeue the according one from $\Omega$
		\STATE $TC_I^N\to TC_U^N$
		\ELSE
		\STATE Reload assignment using the according element in $\Omega$
		\STATE Dequeue the selected element from $\Psi$
		\ENDIF
		\ENDFOR
		\STATE Set all the cells of $\Omega$ in $X$ 1, while the rest 0
	\end{algorithmic}
\end{algorithm}

We take Figure \ref{fig:structure} as an example. The output of CNN 2 for request 2 is $[0.2,0,0.8,0]$ which means the CNN is more confident to cache the content of the second request in edge cloud 3 whose value is $0.8$ compared with caching in edge cloud 1. The size of matrix in PEL is $4\times 3$ indicating we select the proper cache hosts among $4$ edge clouds for $3$ requests. The initial assignment is caching the first content in edge cloud 1 and the rest two in edge cloud 3 with prediction confidence $1$, $0.8$ and $0.6$ respectively, shown with a bold font at the first matrix of PEL. Then, we enumerate the normal element in a descending order until all non-zero element are visited.

\section{Numerical Investigations}
\label{sec:investigations}
In order to compare the performance of the optimal solution derived from the MILP and the proposed CNN based method, we introduce a Greedy Caching Algorithm (GCA) and a Randomized Greedy Caching (RGC) algorithm. GCA attempts to allocate each request to its nearest EC, while the RGC starts from the GCA assignment and then performs random local search by exploring adjacent edge clouds. In that case, when a new assignment with smaller $TC^N$ is found, it  replace the original solution. We assume a mesh tree-like mobile network where user mobility is take placing on the edge of the network. Based on user mobility patterns  we apply the different techniques, i.e., MILP, CNN, GCA and RGC respectively, to efficiently perform pro-active edge cloud caching of popular content. The mobility of users between the different AR $a$ depends on the moving probability $P_{ka}$ and their requests for content. The simulation parameters that assumed in the investigations are summarized in Table \ref{tab:Network_Parameters}. 

\begin{table}[htb]
	\centering
	\caption{\label{tab:Network_Parameters}Network Parameters Used in This Paper\cite{wang2019proactive}.}
	\begin{tabular}{l|c}
		\hline
		\textbf{Parameter}&\textbf{value} \\
		\hline
		number of mobile users ($|\mathcal{K}|$)& \{5,10,15,20\} \\
		weight of cache host ($\alpha$)& [0,1] \\
		weight of transmission cost ($\beta$)& [0,1] \\
		factor of invalid constraint penalty ($\gamma$)& 20 \\
		threshold of prediction probability ($\delta$) & 0.001 \\
		size of user request content ($s_k$) & [10,50]MB \\
		available cache size in EC ($w_e$) & [100,500]MB\\
		user request transmission bandwidth ($b_k$) & [1,10]Mbps\\
		link available capacity ($c_l$) & [50,100]Mbps\\
		number of epochs for RGC & 500 \\
		\hline
	\end{tabular}
\end{table}

The data set is generated by solving the optimization problem \eqref{LP:main} with different user behaviour and network utilization levels, i.e., $p_{ka}$, $s_k$, $b_k$, $w_e$ and $c_l$. In the results presented hereafter, firstly we assume 1000 samples for the scenario of 5 users; out of those $900$ are used for training by using the structure of each CNN illustrated in Figure \ref{fig:cnn} and the rest are using for testing. Then we construct each $100$ samples for 10, 15 and 20 user cases respectively. In order to test the performance on these scenarios, we split grayscale images into some sub images which match the input size of trained CNN, use CNN to allocate caching for this batch and update the rest sub images depending on the preceding allocation recursively.  

In contrast to the normal multi-labels classification problem, some misclassifications are still acceptable if the total cost $TC^N$ is sufficiently close to the optimal result. Firstly, we compare the computation time of these algorithms. For the CNN, this is the time from reading the test samples to getting the output of PEL excluding the training time. Then the average total cost of these approaches are juxtaposed. Next, we compare the final precision of the proposed scheme which is defined as the ratio of mobile users which are predicted correctly in terms of their edge cloud association. After that the feasible ratio of these algorithm are investigated which is the percentage of the constraints-satisfied assignments. Finally, the maximum total cost difference among these algorithms are analyzed, which represents the performance gap in the worst case. The simulations are conducted using MATLAB in a 64-bit Windows 10 environment on a machine equipped with an Intel Core i7-7700 CPU 3.60 GHz Processor and 16 GB RAM. 

The performances are compared in Table \ref{tab:Performance_Comparison}. In these cases, solving the MILP model directly outperforms the others algorithms in terms of mean total cost, allocation precision and conditions feasible ratio. Note that when the network expands to 10 requests, solving the integer mathematical formulation requires $100$ seconds, which cannot be considered practical for real-time decision making. CNN performs similarly to the MILP with less computational time. Even in the 15- or 20-requests case, CNN produces a highly competitive solution. CNN benefits from the numerous training data, which can be viewed as a form of prior knowledge of caching assignments. The GCA has the largest cost payment from all schemes. This is expected by the fact that GCA only consider caching the content to the nearest EC with the aim to reduce the routing cost. But it has the risk to pay a large cost for storage because some edge clouds might be heavily utilized.
RGC tries different assignments on each iteration to find a better solution
but the searching exploration is inefficient due to the inherent complexity of the problem. Observing that the average precision of GCA is higher than RGC in these scenarios, this is reasonable because the updating process of RGC is towards the direction which cause less total cost. 
As can be seen from the table, the gap of mean cost between CNN and MILP becomes larger with the increment of the number of requests but the average value of CNN is still within the two times optimal solution. Furthermore, CNN can provide better solution than RGC, with even more than 400\% increment on performance in the case of $15$ requests (the maximum difference for CNN is $106.6$ but RGC is $463.9$) and more than 50\% improvement on precision.

\begin{table*}[htb]
	\centering
	\caption{\label{tab:Performance_Comparison}Performance Comparison.}
	\begin{tabular}{l|c|c|c|c|c|c|c|c}
		\hline
		&\multicolumn{4}{|c}{$5$ requests}
		&\multicolumn{4}{|c}{$10$ requests}\\
		&\multicolumn{4}{|c}{number of variables:376,number of constraints:782}&\multicolumn{4}{|c}{number of variables:746, number of constraints:1532}\\
		\hline
		&\textbf{MILP}&\textbf{CNN}&\textbf{GCA}&\textbf{RGC}&\textbf{MILP}&\textbf{CNN}&\textbf{GCA}&\textbf{RGC}\\
		\hline
		Computation Time& $0.49s$ & $0.15s$ & $0.01s$ & $9.18s$ & $80.96s$ & $0.51s$ & $0.03s$ & $15.27s$\\
		Mean Total Cost& $10.83$ & $10.84$ & $17.95$ & $16.74$ & $23.31$ & $25.07$ & $58.28$ & $37.09$\\
		Precision& $100.0\%$ & $92.8\%$ & $63.5\%$ & $31.8\%$ & $100.0\%$ & $81.5\%$ & $57.1\%$ & $34.1\%$\\
		Feasible Ratio& $100.0\%$ & $100.0\%$ & $99.8\%$ & $100.0\%$ & $100.0\%$ & $99.3\%$ & $95.2\%$ & $97.9\%$\\
		Maximum Diff& $0$ & $0.4$ & $67.2$ & $37.4$ & $0$ & $89.01$ & $329.0$ & $101.4$\\
		\hline
		\hline
		&\multicolumn{4}{|c}{$15$ requests}&\multicolumn{4}{|c}{$20$ requests}\\
		&\multicolumn{4}{|c}{number of variables:1116,number of constraints:2282}&\multicolumn{4}{|c}{number of variables:1486, number of constraints:3032}\\
		\hline
		&\textbf{MILP}&\textbf{CNN}&\textbf{GCA}&\textbf{RGC}&\textbf{MILP}&\textbf{CNN}&\textbf{GCA}&\textbf{RGC}\\
		\hline
		Computation Time& $1.58h$ & $1.07s$ & $0.04s$ & $16.75s$ & $2.4h$ & $2.1s$ & $0.08s$ & $47.26s$\\
		Mean Total Cost& $39.04$ & $55.59$ & $124.75$ & $98.72$ & $64.91$ & $124.66$ & $269.46$ & $156.36$\\
		Precision& $100.0\%$ & $65.4\%$ & $51.9\%$ & $26.4\%$ & $100.0\%$ & $51.9\%$ & $44.4\%$ & $21.1\%$\\
		Feasible Ratio& $100.0\%$ & $94.8\%$ & $89.0\%$ & $89.8\%$ & $100.0\%$ & $84.9\%$ & $80.8\%$ & $80.8\%$\\
		Maximum Diff& $0$ & $106.6$ & $463.9$ & $463.0$ & $0$ & $136.6$ & $646.1$ & $635.8$\\
		\hline
	\end{tabular}
\end{table*}

\section{Conclusions}
\label{sec:conclusions}
Convolutional neural networks (CNNs), a class of artificial neural networks, have shown remarkable performance in most computer vision tasks.
Inspired by these results in this paper we present an approach of transforming an optimization problem related to  caching of popular content on a number of edge clouds to a grayscale image so that to train deep convolutional neural networks.
Numerical investigations reveal that the proposed scheme can provide  real-time decision making which can be even more than 400\% better than powerful randomized greedy heuristics which are one of the options for real time decision making.  
Future avenues of research are multifaceted. 
Further analysis is needed in the area of sensitivity of the deep learning output to controlled small variations of the incoming requests. This can be viewed as sensitivity analysis as with respect to the integer linear programming formulation and adversarial behaviour of the CNN. Furthermore, an interesting extension would be to include temporal characteristics of the problem that will require new ways of transforming the spatio-temporal aspects of the optimization problem to an image.

\addtolength{\textheight}{-12cm}   

\bibliographystyle{ieeetr}
\bibliography{reference}

\end{document}